\documentclass[useAMS,usenatbib]{mn2e}
\usepackage{amsmath,amssymb,amsfonts,mathrsfs,latexsym}
\usepackage{graphicx}
\usepackage{hyperref}
\begin{document}

\title[Superburst in the NS EXO 1745--248]{A superburst candidate in
  EXO 1745--248 as a challenge to thermonuclear ignition models}

\author[Altamirano et
  al.]{D. Altamirano$^1$\thanks{E-mail:d.altamirano@uva.nl},
  L. Keek$^2$, A. Cumming$^3$, G. R. Sivakoff$^4$, C.O. Heinke$^4$,
  \newauthor R. Wijnands$^1$, N. Degenaar$^5$, J. Homan$^6$ and
  D. Pooley$^{7,8}$ \vspace{0.5cm} \\
  $^{1}$: Astronomical Institute, ``Anton Pannekoek'', University of
  Amsterdam, Science Park 904, 1098XH, Amsterdam, The Netherlands.\\
  $^2$: National Superconducting Cyclotron Laboratory, Department of
  Physics \& Astronomy, and Joint Institute for Nuclear Astrophysics,\\
  Michigan State University, East Lansing, MI 48824, USA.\\
  $^3$: Department of Physics, McGill University, 3600 rue University,
  Montreal, QC, H3A 2T8, Canada.\\
  $^4$: Dept. of Physics, U. of Alberta, CCIS 4-183, Edmonton, AB T6G
  2E1. Canada.\\
  $^5$: Department of Astronomy, University of Michigan, 500 Church
  Street, Ann Arbor, MI 48109, USA.\\
  $^6$: Massachusetts Institute of Technology - Kavli Institute for
  Astrophysics and Space Research, Cambridge, MA 02139, USA.\\
  $^7$: Sam Houston State University, USA. \\
  $^8$: Eureka Scientific, Inc., USA.}

\pagerange{\pageref{firstpage}--\pageref{lastpage}}
\pubyear{2012}

\maketitle

\label{firstpage}

\begin{abstract}

We report on Chandra, RXTE, Swift/BAT and MAXI observations of a
$\sim$1 day X-ray flare and subsequent outburst of a transient X-ray
source observed in October--November 2011 in the globular cluster
Terzan~5.
We show that the source is the same as the transient that was active
in 2000, i.e., the neutron star low-mass X-ray binary EXO~1745--248.
For the X-ray flare we estimate a 6--11 hr exponential decay time and
a radiated energy of $2-9 \times 10^{42}$ erg. These properties,
together with strong evidence of decreasing blackbody temperature
during the flare decay, are fully consistent with what is expected for
a thermonuclear superburst.
We use the most recent superburst models and estimate an ignition
column depth of $\approx 10^{12}$ g cm$^{-2}$ and an energy release
between $0.1-2 \times 10^{18}$ erg g$^{-1}$, also consistent with
expected superburst values.
We conclude therefore that the flare was most probably a superburst.
We discuss our results in the context of theoretical models and find
that even when assuming a few days of low level accretion before the
superburst onset (which is more than what is suggested by the data),
the observations of this superburst are very challenging for current
superburst ignition models.
\end{abstract}


\section{Introduction}
\label{sec:intro}

%
%

Thermonuclear Type-I X-ray bursts are caused by unstable burning of a
several meters thick layer of accreted H/He on the surface of neutron
stars (NSs) in low-mass X-ray binary (LMXB) systems
\cite[e.g.][]{Lewin93}.
Manifesting themselves as a sudden (seconds) increase in the X-ray
luminosity and reaching levels that can be many times brighter than
the persistent (accretion) luminosity,
typical bursts emit about $10^{39}-10^{40}$ ergs, last seconds to
minutes, and have light curves that are well described by a fast-rise
exponential-decay profile.
Their spectra are generally consistent with a blackbody temperature
$T_{\rm bb}=2$--3~keV, where $T_{\rm bb}$ increases until the burst
peak, and then decreases exponentially.
This is naturally interpreted as heating resulting from the initial
fuel ignition, 
followed by cooling of the ashes \citep[and additional hydrogen
  burning through a series of rapid proton captures and
  $\beta$-decays, 
e.g.,][]{Schatz01} once the main available fuel is exhausted.
Type-I X-ray bursts are a common phenomenon in NS-LMXBs. They have been
observed in about 100 sources and, depending on the conditions (e.g,
accretion rate, composition of the fuel, etc.) can have recurrence
times between minutes and weeks \cite[e.g.][]{Galloway08,Linares12}.

Superbursts are a class of extremely long-duration bursts which are
attributed to the unstable thermonuclear burning of a $\sim$100 meter
thick carbon-rich layer, formed from the ashes of normal Type-I X-ray
bursts \citep{Cumming01a}.
Superbursts tend to quench the regular Type-I bursts for weeks
afterwards, probably because the cooling flux from the superburst
temporarily stabilizes the H/He burning
\citep{Cumming01a,Cumming04,Keek12}.
The difference in fuel composition between Type-I X-ray bursts and
superbursts leads to a clear difference in time scales, recurrence
times and energetics, where superbursts last for a few hours, recur
every one-to-few years and emit $10^{41}-10^{42}$ ergs.
With such long recurrence times superbursts are difficult to
catch. While thousands of Type-I X-ray bursts have been observed
\citep[e.g.][]{Galloway08}, to date only about 22 (candidate)
superbursts have been observed from 13 sources \citep[see, e.g., ][and
  references therein]{Wijnands01, Kuulkers04, Keek11, Altamirano11f,
  Chenevez11, Mihara11, Asada11}.

%
%

Terzan 5 is a globular cluster containing ~50 known X-ray sources, of
which $\sim$12 are likely LMXBs containing neutron stars
\citep[e.g.][]{Heinke06}.
During 2011 we monitored Terzan 5 on a weekly basis with \textit{Rossi
  X-ray Timing Explorer} (RXTE) observations to search for transient
X-ray flares and/or outbursts.
At 4:57 UT, October 26th 2011, an RXTE pointed observation measured a
2--16 keV intensity of $\sim$8 mCrab, significantly above the typical
quiescent intensity of $\sim$2 mCrab \citep{Altamirano11e}.
Approximately 8 hours earlier, INTEGRAL monitoring observations of
Terzan 5 did not detect any enhanced activity, with a $5\sigma$ upper
limit of 6 mCrab in the 3--10 keV energy band \citep{Vovk11}.
The RXTE detection was confirmed by the Swift/BAT daily-averaged flux
measurements \citep{Altamirano11e}, as well as by a Swift/XRT pointed
observation performed $\sim$11 hours after the RXTE one
\citep{Altamirano11f}.
The position of the source from these Swift/XRT data was consistent
\citep{Altamirano11f,Evans11} with that of the transient NS-LMXB that
was active in 2000 \citep[which we refer to as EXO 1745--248, though
  it is not necessarily the EXOSAT source; see][]{Markwardt00,
  Wijnands05a}. This result was later confirmed by a preliminary
analysis of a pointed Chandra observation \citep{Pooley11}.

Just before the INTEGRAL non-detection, MAXI and Swift/BAT light
curves of Terzan 5 revealed an X-ray flare that lasted less than a
day.
We identified this flare as a possible superburst based on its
duration, shape of its light curve and estimated radiated energy of
$\sim$10$^{42}$ ergs \citep{Altamirano11f}.
Our speculations were supported by the results of \citet{Mihara11} who
used the MAXI data and showed that (i) the spectra of the flare were
well modeled with a blackbody component at $\sim$2--3~keV and that
(ii) there was an apparent decrease of the black-body temperature,
which is usually interpreted as the cooling of the neutron star
surface after a thermonuclear burst \citep[see,
  e.g.,][]{Lewin96}. Very recently, \citet{Serino12} have presented a
detailed analysis of the MAXI data supporting the superburst
identification.
The occurrence of a superburst in the transient NS-LMXB 4U~1608--522
after 55 days of low ($\lesssim10$\% Eddington) accretion rate has
challenged superburst theory, as it is difficult to explain carbon
ignition at the observed depths when the NS surface is still cool
\citep{Keek08}, i.e., when accretion has not yet been able to ``warm
up'' the NS.
The superburst candidate in Terzan 5 is even more challenging for
theoretical models, as the NS is very cool \citep{Wijnands05a,
  Degenaar12} and the superburst onset was coincident with a period of
only low-level accretion, or no accretion at all.

\begin{figure}
\centering
\resizebox{1\columnwidth}{!}{\rotatebox{0}{\includegraphics{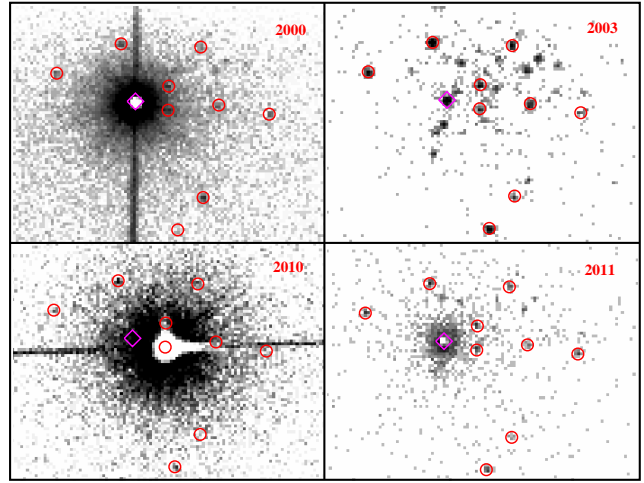}}}
\caption{39" x 52" Chandra images of Terzan 5 from different epochs
  show the 2011 outburst source is EXO 1745--248.
The upper-left panel shows the combined image of two observations
(July 24 and 29, 2000, ObsIDs 655 and 644, respectively) for a total
time of 47 ksec \citep[see][]{Heinke03a}.  The upper-right panel shows
a 35-ksec observation on July 13th, 2003 \citep[ObsID 3798;
  see][]{Wijnands05a,Heinke06}.  The lower left panel is a 10 ksec
observation on Oct. 24, 2010 \citep[ObsID 11051;][]{Pooley10}, and the
lower right panel is our 9.8 ksec observation on Nov. 3, 2011
\citep[ObsID 12454;][]{Pooley11}.  All images were extracted in the
1-3 keV energy range (chosen to try to maximize S/N in the 2011
image).
 Ten X-ray sources from \citet{Heinke06} are marked with red circles
 (2-pixel -- 0.984" radius). Diamonds mark the position of
 EXO~1745--248 as detected in its 2000 outburst. The active source in
 the 2010 observation \citep{Pooley10} was the 11~Hz pulsar
 IGR~J17480--2446 \citep{Strohmayer10a,Papitto11} }\label{fig:chandra}
\end{figure}

\section{OBSERVATIONS AND DATA ANALYSIS}
\label{sec:dataanalysis}

In this paper we make use of data from the MAXI \citep{Matsuoka09},
Swift/BAT \citep[Burst Alert Telescope][]{Barthelmy05}, Chandra
\citep{Garmire03}, INTEGRAL \citep{Winkler03} and RXTE
\citep{Jahoda96} missions. 
Most data presented here were obtained during October-November
2011. However, as we explain below, we also used archival data sets
from different periods to put our results in a long-term context.

We used the processed MAXI data as provided by the MAXI Team: four
light curves are available (corresponding to the 2--4 keV, 4--10 keV,
10-20 keV and 2-20 keV energy bands), which are given in either 1-day
or 1 orbit bins\footnote{http://maxi.riken.jp/top/}.
We also used data from the Swift/BAT transient
monitor\footnote{http://swift.gsfc.nasa.gov/docs/swift/results/transients/}.
These 15--50 keV data are provided by the Swift/BAT team after being
processed, corrected for systematic errors and binned in both daily
and orbital bins.

To identify the active source in Terzan 5, we obtained a 9.8 ks
Chandra observation (ObsID 12454, November 3rd, 2011 at 5:05:57 UTC.)
taken with the ACIS S3 chip in imaging mode.
We also used other available Chandra observations of Terzan 5
\citep[e.g.,][]{Wijnands05a, Heinke06, Pooley10}.
All images were reprocessed with CIAO 4.3 following standard
recipes\footnote{http://cxc.harvard.edu/ciao/threads/}.

We used 18 pointed observations of the RXTE Proportional Counter Array
\citep[PCA; for instrument information see][]{Zhang93,Jahoda06} that
sampled the 2011 outburst in Terzan~5.
We use the 16-s time-resolution Standard~2 mode data to calculate the
Crab-normalized 2.0--16.0 keV intensity as described in
\citet{Altamirano08}.
For the timing analysis we used Event mode E\_125us\_64M\_0\_1s or the
Good Xenon data. Power spectra were generated following
\citet{Altamirano08} using data segments of 128 seconds and 1/8192~s
time bins.
To fit the power spectra, we used a multi-Lorentzian function.
We only include those Lorentzians in the fits whose single trial
significance exceeds $3\sigma$ based on the error in the power
integrated from 0 to $\infty$ and we give their frequency in terms of
characteristic frequency \citep{Belloni02}.
The quoted errors use $\Delta\chi^2 = 1.0$ corresponding to a 68\%
confidence level.

\begin{figure*}
\centering
\resizebox{1\columnwidth}{!}{\rotatebox{0}{\includegraphics{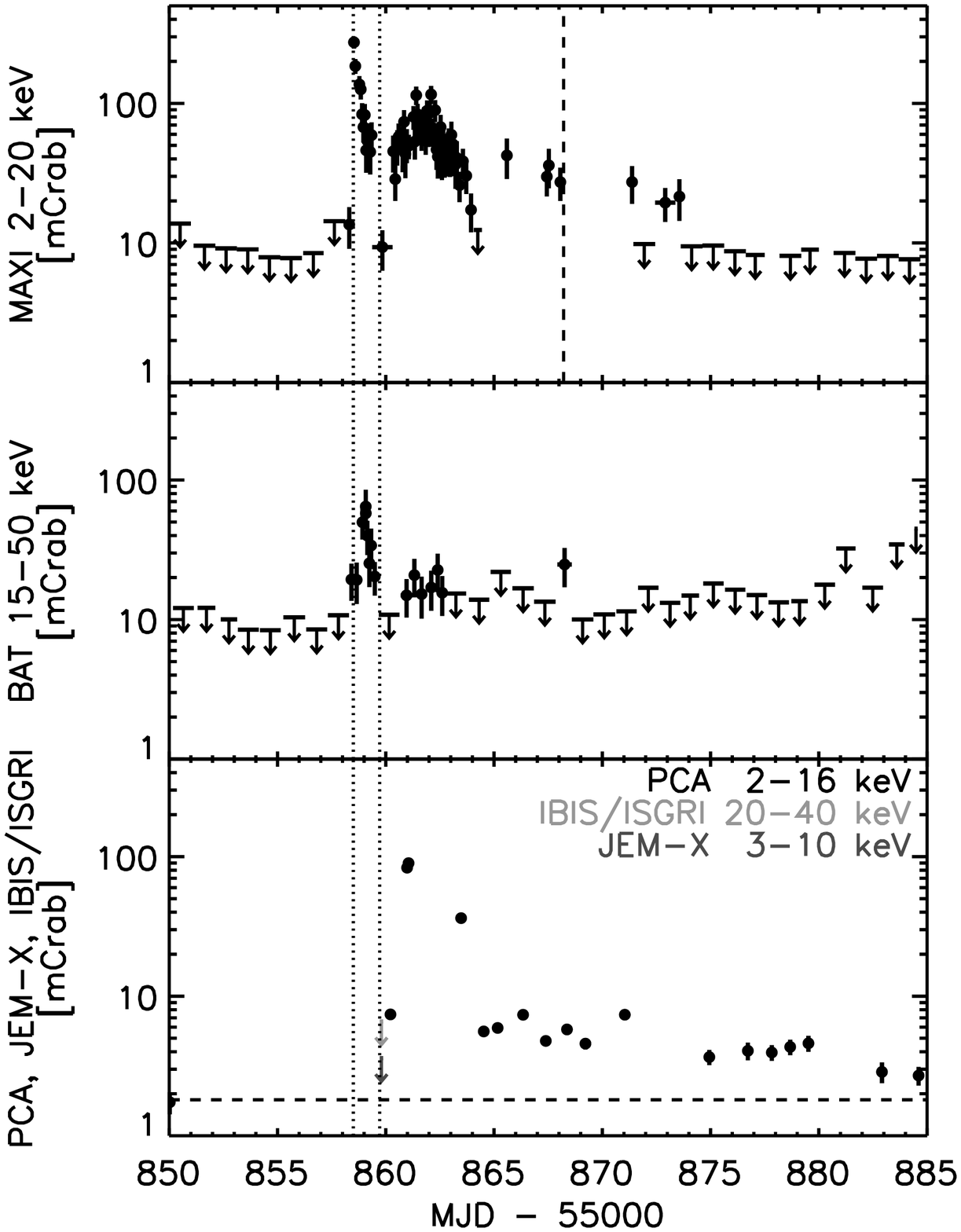}}}
\resizebox{1\columnwidth}{!}{\rotatebox{0}{\includegraphics{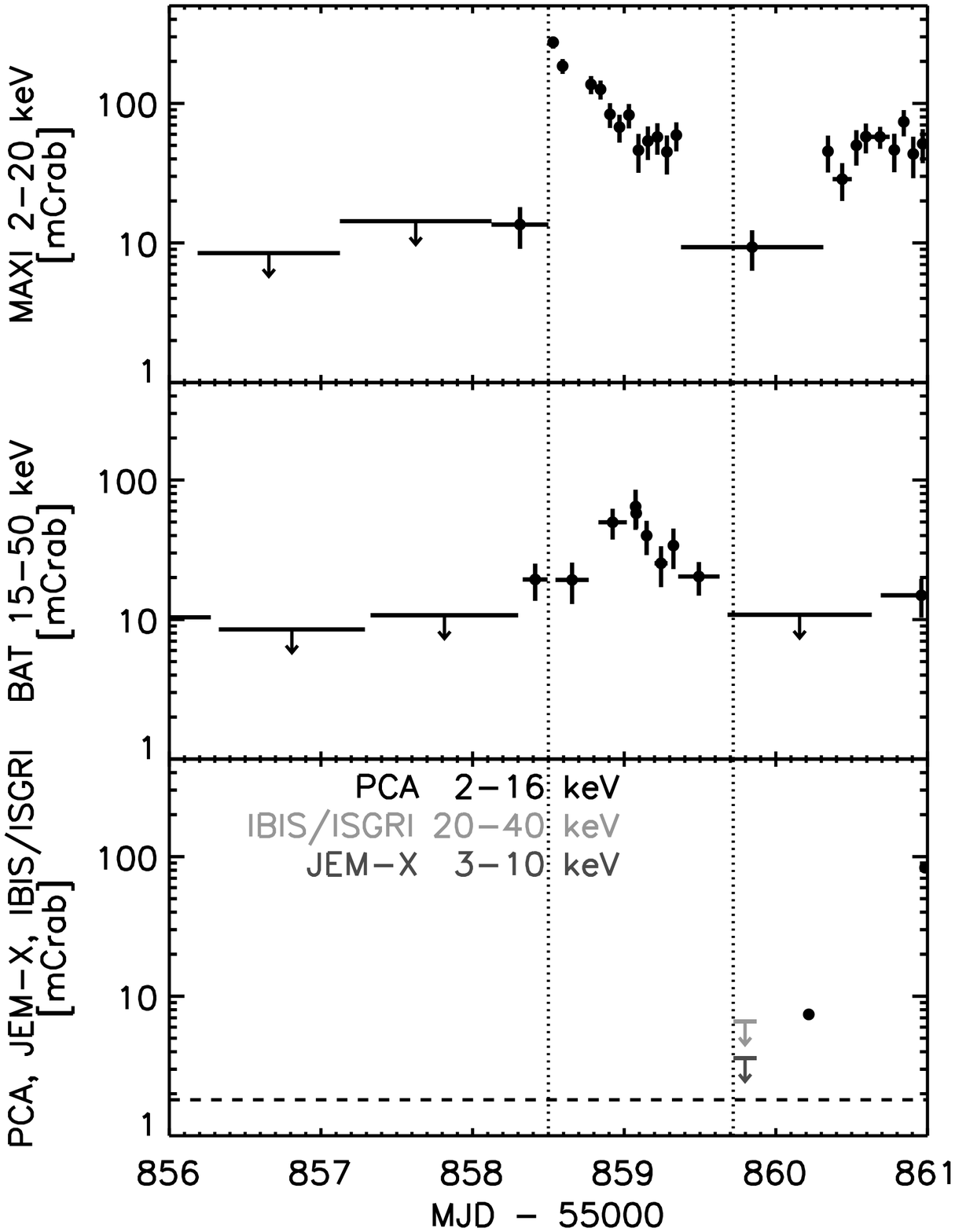}}}
\caption{X-ray light curves of EXO~1745--248 in Terzan 5 as sampled
  with MAXI (upper panels), Swift/BAT (middle panels) and RXTE pointed
  observations (lower panels).
For MAXI we subtracted an average background of 0.05 cts/sec before
converting to Crab. MAXI and Swift/BAT data points are either a
3$\sigma$ detection or a 3$\sigma$ upper limit (see
Section~\ref{sec:lc} for more details). The dashed-line in the
upper-left panel marks the time of the Chandra observation (MJD
55868). The arrows in the lower panels marks the time of the INTEGRAL
2--10 keV 6~mCrab upper limits (with IBIS/ISGRI upper limit higher
than JEM-X) while the circles mark the average per observation RXTE
2--16 keV intensity. The horizontal dashed line in the lower panels
marks the average background emission as estimated from 10 months of
RXTE non-detections before the 2011 outburst in Terzan~5. These values
can be taken as upper limits to the intensity.
Vertical lines mark the approximate region between the onset and end
time of the superburst. Right panels show zoom-ins to this
region.}\label{fig:lc}
\end{figure*}

Recent estimates of the distance to the globular cluster Terzan 5
range between 4.6 kpc and 8.7 kpc \citep{Cohn02, Ortolani07,
  Lanzoni10}.  
The large distance range to this globular cluster is mainly
  due to an ongoing discussion on how to identify the horizontal
  branch in the Hertzsprung-Russell diagram of Terzan 5 and the
  assumed reddening factor in the direction of this globular
  cluster. We refer the reader to the discussion reported in
  \citet{Ortolani07}. Following these authors' discussion, in this
  paper we use a distance of $5.5\pm0.9$ kpc which falls in-between
  the different estimates and comes from Hubble Space Telescope
  photometry of Terzan 5 \citep{Ortolani07}.

\section{Results}\label{sec:results}

\subsection{Identification of the source in Terzan 5}

Inspection of the 2011 Chandra data reveals a bright transient (which
suffers from pile up) and a few low-luminosity X-ray sources that are
also seen in deeper Chandra observations in quiescence, such as the
2003 observation shown in Figure~\ref{fig:chandra}. We match 10 of the
brighter 2011 sources with sources in the 2003 observation, allowing
us to confidently identify (within 0.2") the 2011 X-ray transient in
Terzan 5 with CXOGlb J174805.2-244647 (CX3 in \citealt{Heinke06}, at
J2000 coordinates 17:48:05.236 (0.002), -24:46:47.38 (0.02)).
In Figure~\ref{fig:chandra} we show Chandra images of Terzan 5 from
different epochs including that of 2011.
Given that the 2000 outburst was identified with EXO~1745--248
\citep{Markwardt00, Wijnands05a}, in the rest of this paper we refer
to the source as EXO~1745--248.

\subsection{Long term light curves} \label{sec:lc}

In the left panel of Figure~\ref{fig:lc} we show the 2--20 keV MAXI
(upper panel), 15-50 keV Swift/BAT (middle panel) and 2--16 keV RXTE
light curve (lower panel). In the lower panel we also show the
INTEGRAL upper limits; in the upper panel we mark the time of the
Chandra observation. Right panels show a zoom-in to the moment of the
initial X-ray flare.
Due to the low statistics of both the MAXI and Swift/BAT orbital data,
we used an adaptive binning method which (i) has been fixed to start
around the beginning of the flare (MJD 55859.5) and (ii) bins
observations until finding a 3$\sigma$ detection within a day, or
calculates a 3$\sigma$ upper limit for 1 day of data.
Figure~\ref{fig:lc} shows that the peak of the X-ray flare as seen by
MAXI occurs approximately half a day before that of Swift/BAT. Other
binning methods led to similar results.
Figure~\ref{fig:lc} also shows that both MAXI and Swift/BAT appear to
have detected the source before the peak of the flare in the MAXI
data. In a period of $\sim$250 days before the flare we find five
similarly significant detections in the MAXI light curve and two in
the Swift/BAT light curve. These events did not occur simultaneously
in MAXI and Swift/BAT and even if real, cannot be
  unambiguously identified with EXO~1745--248 due to the large number
  of X-ray sources in Terzan 5 \citep[e.g.,][]{Heinke06}.
The fact that we find simultaneous excesses in the MAXI and the
Swift/BAT data is suggestive of a real increase of flux before the
flare; however, given the lack of further information we decided to
take these detections only as marginally significant given the
systematic errors and possible background issues which we are unable
to account for.

The last RXTE/PCA observation was performed on November 19th, 2011,
after which the source was not visible anymore due to visibility
constraints. The lower-left panel of Figure~\ref{fig:lc} shows that
the outburst lasted at least 25 days (still ongoing at the moment of
the last pointed observation); but that the source was brighter than
10 mCrab for only about 4 days.

Terzan 5 was not visible to X-ray instruments (due to Sun constraints)
for the next couple of months.  The next pointed observation of Terzan
5 was on Feb. 9, 2012, for 972 seconds with Swift-XRT, which showed a
count rate of 0.015 cts/s, translating to a total
$L_X\sim6\times10^{33}$ ergs s$^{-1}$.  As this is consistent with the
typical integrated X-ray luminosity of the cluster sources in
quiescence, we conclude that the outburst was finished by then, having
lasted between 25 and 106 days.

\subsection{Type-I X-ray bursts and Superbursts}\label{sec:bol}

We searched all RXTE observations of Terzan 5 that were taken
in 2011 (up until November 19th) for Type-I X-ray bursts, but
none were found.
Lower limits on X-ray burst recurrence times are unconstrained, as our
data set consists of about 27 hr of data in about 15 days, i.e., at an
average of less than 2 hr a day.
The MAXI data consist of about 100 orbital data sets with an average
length of less than a minute each. None of these pointings show
evidence for an X-ray burst \citep[see also][]{Serino12}.

To calculate the bolometric luminosity, radiated energy, and e-folding
timescale of the flare we used the background-corrected 2--4 keV MAXI
data during the period MJD 55858.5-55859.5. (We did not use the 2--20
keV lightcurve to avoid systematics related to the flux conversions
between a 2 keV blackbody and a 2.1-index powerlaw as in the Crab
Nebula spectra).
We estimated the RXTE Crab flux in the 2--4 keV range to be $1.0326
\times 10^{-8}$ ergs cm$^{-2}$ s$^{-1}$. For the 2--4 keV MAXI light
curve\footnote{see http://http://maxi.riken.jp/top/}, 1 Crab equals
1.87 photons cm$^{-2}$ s$^{-1}$.
With the above values we transformed the 2--4 keV intensity (photons
cm$^{-2}$ s$^{-1}$) into flux.
We then followed \citet{Mihara11}\footnote{The method used by
    \citet{Mihara11} is instrument dependent, and consists of the
    estimation of hardness ratios for given kT parameters through
    simulated energy spectra using the MAXI energy response. The
    energy ranges used in this paper and in \citet{Mihara11} allows
    estimates of the temperature kT from the hardness ratios
    (M. Serino, private communication). Similar methods have been used
    for RXTE (e.g., Figure 1 in \citealt{Belloni00}).}
 and approximated blackbody color temperatures from the 4--10 keV/
 2--4 keV color hardness.
Then we used
PIMMS\footnote{http://heasarc.nasa.gov/Tools/w3pimms.html}, the
absorbed flux, and color temperatures estimated above to approximate
the unabsorbed \citep[assuming Galactic $N_H=1.2 \times 10^{22}$
  cm$^{-2}$,][]{Altamirano11f} bolometric flux of a black body in the
0.01-200 keV range.
We finally converted our values to bolometric luminosities assuming a
distance of 5.5 kpc \citep{Ortolani07}.
The bolometric luminosity of the superburst as a function of time is
shown in Figure~\ref{fig:sim}.

Due to: (i) all of the assumptions made to calculate the bolometric
luminosity, (ii) the fact that the MAXI data does not sample the
beginning of the flare (which could have happened at any time in the
90 min between orbits), and, (iii) the fact that under the hypothesis
that the flare is from a thermonuclear origin, we assume the
contribution from the accretion disk to be negligible, it is not
possible to get tight constraints on the characteristics of the flare.
Since the first MAXI data point sampling the flare is at $\sim6 \times
10^{37}$ (D/5.5 kpc)$^2$ ergs s$^{-1}$, the flare peak was probably
brighter.
The flare duration is about a day; an exponential fit to the
bolometric luminosity light curve gives e-folding times between 6 and
11 hr depending on the assumed onset time. Exponential fits to the
raw 2-20 keV MAXI data give consistent results.
Integrating this exponential curve during a 1 day period gives a
radiated energy in the $2-9 \times 10^{42}$ ergs range. More than
$85$\% of the contribution comes from the first 5 hr.
All of these values are within the ranges expected for superbursts,
although this one appears to be one of the longest such bursts
\citep[see, e.g.,][]{Keek08a}.
Our results, together with the fact that the MAXI spectra of the flare
are consistent with blackbody spectra at $\sim$2--3 keV that cool as
the intensity decreases \citep{Mihara11,Serino12}, strongly suggest
that the observed X-ray flare is most probably a superburst.

We note that recently \citet{Serino12} reported on the spectral
modeling of the same MAXI data used in this work.
Their spectral results are binned into 5 intervals (A-E) of different
time-lengths to improve the S/N of their energy spectra.
The method used in this paper to estimate the blackbody temperature
and flux is more rudimentary, but has the advantage of giving more
points for fitting the thermal evolution of a superbursts (see
Section~\ref{sec:fit}). Given that the values of bolometric luminosity
and radiated energy \citet{Serino12} obtain are consistent with ours
within errors (after correcting for the fact that they estimated the
fluence in the 2--20 keV range, and used 8.7 kpc as the distance to
Terzan 5), in the following Sections we use the values for the
bolometric luminosity as we calculated above.

\subsection{Ignition depth and energy release}\label{sec:fit}

\citet{Cumming04} modeled the thermal evolution of the surface layers
as they cool after a superburst onset, assuming that the fuel is
burned locally and instantly.
These authors showed that simultaneous modeling of superburst light
curves and quenching times could be used to constrain both the
thickness of the fuel layer and the energy deposited in the neutron
star envelope.
\citet{Cumming06a} applied the \citet{Cumming04} models to the
observations of several superbursts and found that their fits implied
ignition column depths in the range $(0.5-3) \times 10^{12}$ g
cm$^{-2}$, energy releases on the order of $\approx 2 \times 10^{17}$
ergs g$^{-1}$, and total radiated energies on the order of $10^{42}$
ergs, very similar to the observed superburst characteristics.

The model has four free parameters to vary: energy release $E_{18}
\times 10^{18}$ ergs g$^{-1}$, ignition column depth
$y_{\mathrm{ign}}$ (in units of g cm$^{-2}$), burst start time, and
the power-law slope of the initial temperature profile T$ \propto
y_{\mathrm{ign}}^\alpha$.
The model assumes a 1.4 $M_\odot$ and a 10 km radius NS.
\citet{Cumming06a} assumed that the fuel burned instantaneously
``in~place'', giving an initial temperature profile with
$\alpha\sim$1/8. Instant burning implies that the rise of the burst is
instantaneous. Here we also explore $\alpha=0.225$, which is required
to fit the rise of the superburst light curve of 4U~1636--53 (this fit
and its implications will be published elsewhere).

As is usually the case in the spectral analysis of Type-I X-ray
bursts, it is possible that the tail of the superburst can be
contaminated by the accretion disk.
To understand the possible contribution, we fit all superburst data as
well as data from the first 0.5 days (12 and 6 independent points,
respectively). We assumed a distance of $5.5\pm 0.9$~kpc and allowed
the superburst start time to vary between 0 and 5800 seconds before
the first data point.
To model the superburst lightcurve we used a Markov chain Monte Carlo
method with 30,000 samples.

\begin{figure}
\centering
\resizebox{1\columnwidth}{!}{\rotatebox{0}{\includegraphics{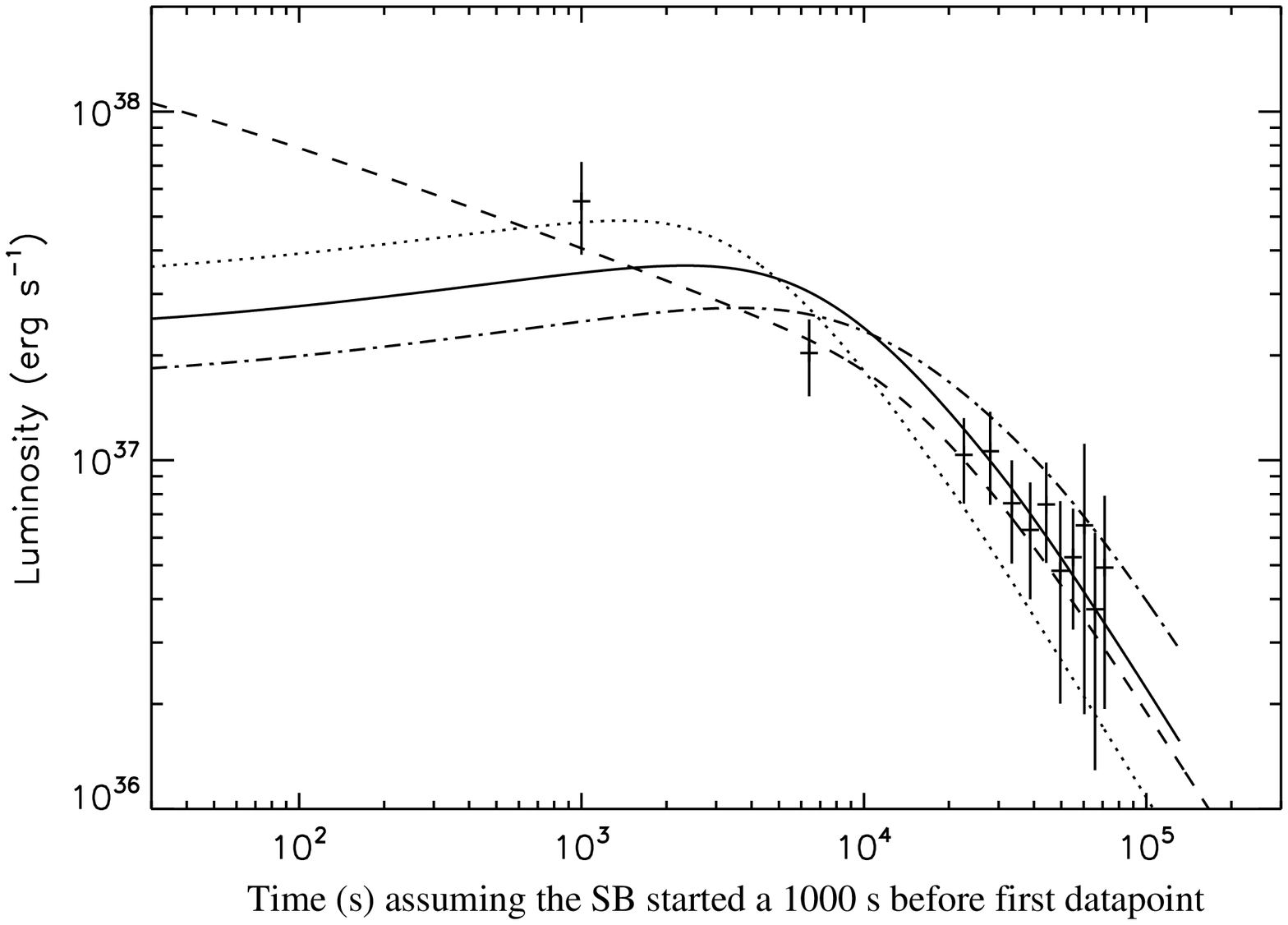}}}
\resizebox{1\columnwidth}{!}{\rotatebox{0}{\includegraphics{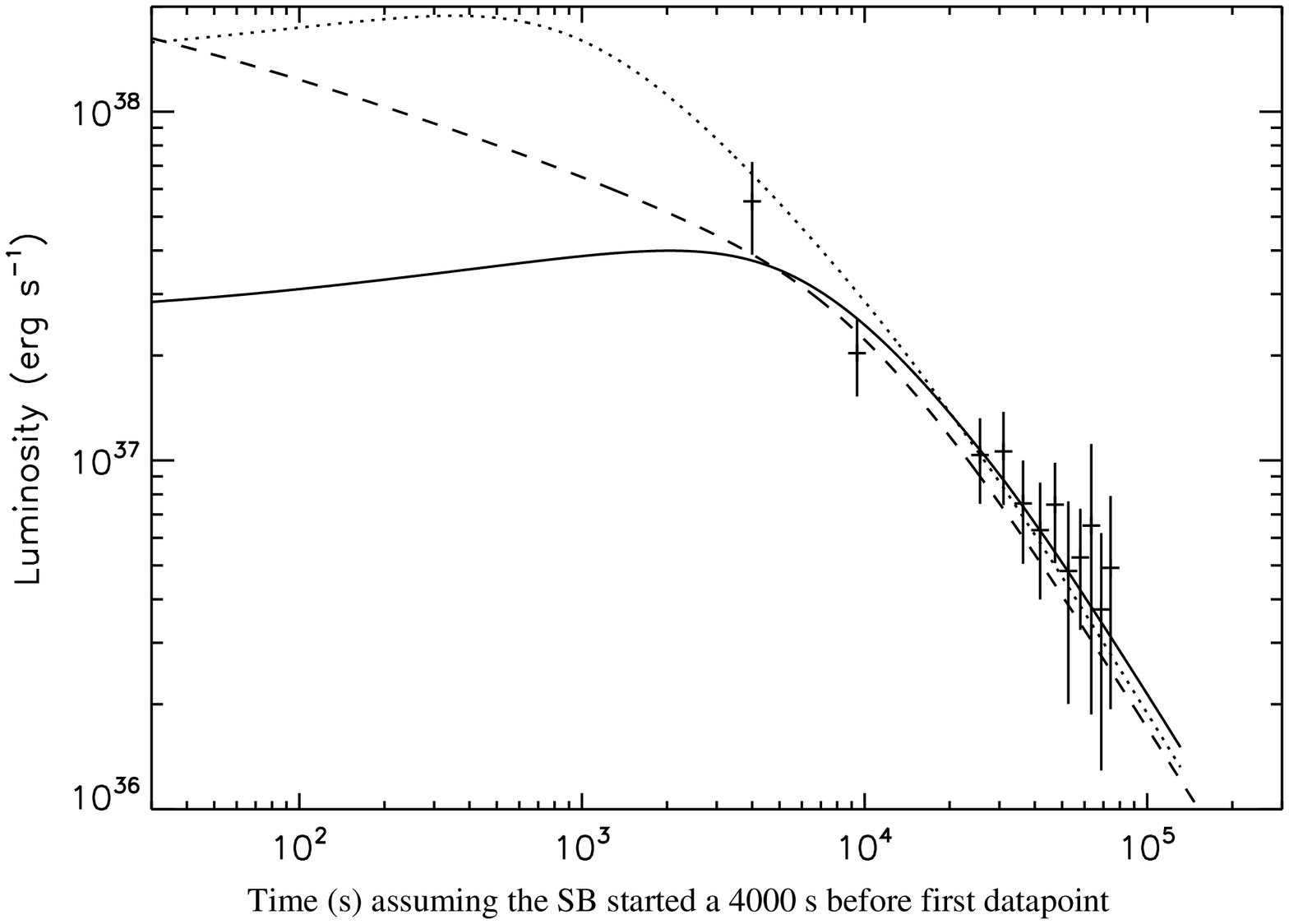}}}
\caption{Upper and lower panel show representative fits to the first
  6 data points assuming MAXI observations starting 1000 s and 4000 s
  after superburst ignition, respectively.
  \textit{Upper panel:} solid, dotted, dot-dashed and dashed curves
  correspond to $E_{18} = 0.28, 0.28, 0.28, 0.175$ and $y_{\mathrm{ign}} =
  12.06, 11.8, 12.3, 12.1$, respectively.
  \textit{Bottom panel:} solid, dotted and dashed curves correspond to
  $E_{18} = 0.29, 0.6, 0.225$ and $y_{\mathrm{ign}} = 12.04, 11.9, 12.0$,
  respectively.
  Solid, dotted and dot-dashed curves assume $\alpha=0.225$ while the
  dashed curves assume $\alpha=1/8$, i.e., instantaneous burning.}\label{fig:sim}
\end{figure}

\begin{figure}
\centering
\resizebox{1\columnwidth}{!}{\rotatebox{0}{\includegraphics{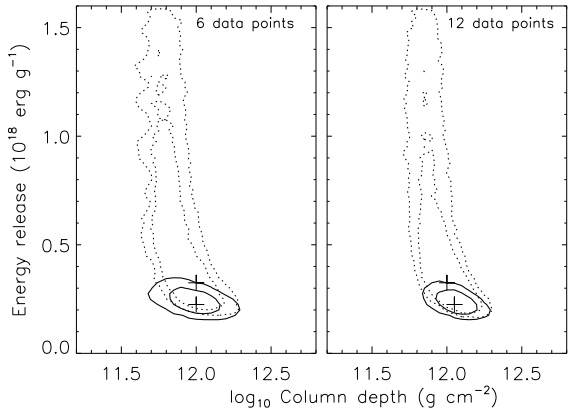}}}
\caption{Contours for $\alpha=0.225$ when using the first 6 data
  points (left) vs 12 data points (right) of the superburst. Filled
  curves are for a superburst starting time of 1000 s while dashed
  curves for 4000 s. Contours are at 68\% and 95\% confidence
  level. }\label{fig:sim1}
\end{figure}

Although we find that $y_{\mathrm{ign}}$ is well constrained to
$\log{y_{\mathrm{ign}}}=12.0\pm0.3$ (including errors in the
distance), our fits failed to constrain the start time and $E_{18}$.
This is due to the fact that our data does not sample the initial rise
of the superburst:
as explained by \citet{Cumming04}, during the first part of the
superburst the energy released from the surface is mainly sensitive to
$E_{18}$ and insensitive to $y_{\mathrm{ign}}$. However, as the superburst
evolves, the characteristics of the cooling tail mainly depend on
$y_{\mathrm{ign}}$.

In the upper and lower panel of Figure~\ref{fig:sim} we show
representative fits to the first 6 data points of the superburst for
different $E_{18}$ and $y_{\mathrm{ign}}$, assuming that the first MAXI
observation occurred 1000 or 4000 seconds after the superburst
ignition, respectively.

For a 1000 s start time (upper panel), our data sample the first hour
of the superburst and therefore the model allows only a narrow range
of $E_{18}$. The column depth is not as well constrained when using 6
points (but it is when using 12 points).
For a start time of 4000 s (bottom panel), however, the first data
point becomes part of the cooling tail. This means that the early part
of the lightcurve can be very bright, $E_{18}$ becomes poorly
constrained while $y_{\mathrm{ign}}$ is well-constrained when using both 6 and
12 points.
In Figure~\ref{fig:sim1} we show the relation between $E_{18}$ and
$y_{\mathrm{ign}}$ ($\alpha=0.225$).  Left and right panels are for
fits to the first 6 and 12 superburst data points, respectively.
We note that a larger distance than that assumed in this paper
  would imply that the data points of the superburst move to higher
  luminosity. This requires a larger energy release to increase the
  luminosity in the early part of the superburst, and a deeper
  ignition depth to increase the luminosity in the cooling tail. For
  the largest distance estimate of 8.7 kpc, we find that the inferred
  energy release increases to close to $10^{18} \ {\rm erg g^{-1}}$
  and the column depth close to $10^{13}\ {\rm g\ cm^{-2}}$.

\subsection{Short (sub-second) variability}

Our power spectral analysis does not reveal major features. In the
first observation of the outburst (MJD 55860.2) we only find a
$3.2\sigma$ (single trial) 500$\pm$20 Hz QPO. In the following three
observations which sample the rest of the bright part of the outburst,
the power spectra are well described by a combination of 3
zero-centered Lorentzians with $\nu_{max}$ at $\sim$0.002, $\sim$1 and
$\sim$15 Hz. After MJD 55864.5 we only detect evidence for power-law
low-frequency noise. Adding all these observations to increase
statistics did not reveal any additional feature.

\section{Discussion}\label{sec:discussion}

We present Chandra, RXTE, Swift/BAT and MAXI data of the X-ray flare
and subsequent outburst of EXO~1745--322 in Terzan~5. We show that the
active source is the same as that active in 2000 and that the
characteristics of the flare are consistent with what is expected for
a superburst.
We also show that the outburst may have started just before the
superburst onset, although our results are not conclusive due to
systematics in the data.
The Swift/BAT peak in the superburst flux was delayed by about 0.5
days compared to the flux peak on the MAXI data.
Similar delays between soft and hard energy bands have already
  been seen in Type-I X-ray bursts \citep[order of seconds, see,
    e.g.,][]{Lewin93,Falanga08,Chelovekov11} and in at least one
  superburst \citep[about $\sim$1000 sec in the LMXB 4U 1820--30, see,
    e.g.,][]{Intzand10}.
These delays have been interpreted as due to photospheric-radius
expansion (PRE) bursts, where the X-ray intensity first peaks in the
low-energy band and later X-rays become visible at higher energies
\citep[see, e.g., ][]{Lewin93,Intzand10}.
The $\sim1000$ sec duration of the PRE phase in the superburst
observed in the LMXB 4U 1820-30 is already at the limit of what
current superburst models can explain.
Irrespective of the mechanism, the delay is by far the largest. The
fact that it is so much larger, may raise the question of its origin
being the same as that proposed for Type-I PRE X-ray bursts.

In Section~\ref{sec:lc} we show marginal evidence that EXO 1745--248
may have been detected before the peak of the superburst. In the rest
of this section we will discuss the implications of our results on
superburst theory taking into account both the possibilities that the
outburst started a few days before, or approximately a day after the
peak of the superburst.
For a discussion on how the superburst emission may have affected the
accretion disk to trigger the subsequent outburst, we refer the reader
to \citet{Serino12}.
We note that if the pre-superburst detections of the source are real,
then the superburst most probably momentarily affected the normal
outburst evolution \citep[see, e.g., ][for the study of the evolution
  of the accretion disk around the NS system 4U 1820--30 during a
  superburst]{Ballantyne04}.

\begin{table}
\begin{centering}
\begin{tabular}{lll}
\hline 
 & EXO~1745--248 & 4U~1254-690\tabularnewline
\hline 
$\tau_{\mathrm{exp}}(\mathrm{hr})$ & $6-11$ & $6\pm0.3$\tabularnewline
$E_{\mathrm{b}}(10^{42}\,\mathrm{erg})$ & $2-9$ & $0.8\pm0.2$\tabularnewline
$\log(y/(\mathrm{g\, cm^{-2}}))$ & $12.0\pm0.3$ & $12.4$\tabularnewline
$E_{18}(10^{18}\,\mathrm{erg\, g^{-1}})$ & $>0.1$ & $0.15$\tabularnewline
\hline 
\end{tabular}
\par\end{centering}\caption{Comparison to the superburst of
  4U~1254-690 \citep{Intzand03,Cumming06a}.}\label{table:table}
\end{table}

\subsection{Comparison to other superbursts and theoretical implications}

Previously, the longest and most energetic superburst known from a
hydrogen accreting source was from 4U 1254--690 \citep{Intzand03a}.
Unlike for that superburst, we did not observe the start of the
superburst from EXO~1745--248, resulting in large uncertainties in the
superburst properties. The superburst of EXO~1745--248 is \emph{at
  least} of equal duration, and twice as energetic
(Table~\ref{table:table}).  The largest values of the bolometric
radiated energy, $E_{\mathrm{b}}$, consistent with the observations,
are close to the predicted maximum radiated energy for a superburst
set by neutrino emission (\citealt{Keek11}; see also
\citealt{Cumming06a}).

The decay time, $\tau_{\mathrm{exp}}$, depends on the thickness of the
cooling layer, and, therefore, on the ignition column depth, $y_{\mathrm{ign}}$.
For 4U~1254--690 the depth was determined using the instantaneous
burning model, yielding a depth comparable to the larger values in the
range we derive for EXO~1745--248, which are also favored by our fits
with the same model \citep{Cumming06a}. This suggests that the
ignition depths and decay times of the two superbursts likely have
similar values. The larger $E_{\mathrm{b}}$ for EXO~1745--248 can be
explained by the burning of more carbon-rich material, which is
accommodated by the larger values in the range found for the specific
energy release, $E_{18}$. Therefore, this is the most energetic and
possibly the longest superburst observed to date.

Most superbursting sources, including 4U~1254--690, are observed to
accrete continuously at a high rate of around $10\%$ of the Eddington
limited rate $\dot{M}_{\mathrm{Edd}}=2\times10^{-8}\,
M_{\odot}\mathrm{yr}^{-1}$ \citep[for solar composition and a
  $10\,\mathrm{km}$ radius; e.g.,][]{Keek08a}.  
The high rate ensures a hot outer crust, forces unstable ignition of
the carbon, and may be necessary for the production of a mixture of
carbon and heavy isotopes that is thought to be the fuel for
superbursts (\citealt{Cumming01a}; see also \citealt{Cooper09}).
Both sufficient heat and carbon are required for superburst ignition.
This scenario was challenged by the observation of the superburst from
the NS-transient 4U~1608--522, which occurred only 55 days after the
onset of an accretion outburst.
 \citet{Keek08} showed that the neutron star envelope does not heat up
 quickly enough to explain the ignition of runaway carbon burning. In
 the past year three more superbursts have been detected from
 transient sources, including the one discussed in this paper
 \citep[for the other detections see][]{Chenevez11,Asada11}.  Even if
 we assume that EXO~1745--248 started accreting at an increased rate
 0.5 days or even a few days before the superburst (but at levels
 undetected by MAXI and Swift/BAT), the time is much too short for the
 envelope at the derived ignition depths to heat up from either
 thermonuclear burning in the envelope or from nuclear processes in
 the inner crust.
Therefore, sufficient heat must have been generated at the superburst
ignition depth within this short time interval. Currently there is no
known process that could provide this. The case for a substantial
additional heat source in the outer crust (close to the superburst
ignition depth) has also been made from observations of crustal
cooling after outbursts in long-duration transients
\citep{Brown09}. The 0.5 days time scale that we find,
however, puts strong constraints on the immediacy with which this
heating process must take place.

\subsection{On the Carbon production}

But where does the carbon-fuel necessary for a superburst come from?
Hydrogen-accreting superbursters display a high ratio of the
persistent fluence between two (Type-I X-ray) bursts to the burst
fluence \citep{Intzand03a}, indicating that apart from during Type-I
X-ray bursts, a substantial fraction of the accreted hydrogen and
helium burns in a stable manner.
This is thought to be a required process to produce the carbon fuel
for superbursters \citep{Schatz03}, and it is observed to occur close
to an accretion rate of $10\%\,\dot{M}_{\mathrm{Edd}}$, i.e., the rate
inferred for most superbursters.
During the outburst in 2000, EXO~1745--248 accreted at a comparable
rate of on average $17\%\,\dot{M}_{\mathrm{Edd}}$ for two months
\citep{Degenaar12}, during which there were bursts as well as periods
without bursts. In fact, because the quiescent luminosity is over a
factor $10^{4-5}$ lower \citep[$L_x$ in quiescence is $5-7 \times
  10^{32}$ erg s$^{-1}$, see][]{Degenaar12}, effectively all of the
superburst fuel must have been created in such short outbursts.
This conclusion is still valid even if we consider that at the above
level of quiescent emission, EXO~1745--248's luminosity might vary by
a factor of a few on timescales of hours-years \citep[which may
  indicate that the accretion does not fully switch off in quiescence,
  but continues at a very low rates, see, e.g.,][]{Wijnands05a}.

During the outburst in 2000, about $8\%$ of the inferred
$y_{\mathrm{ign}}=1.0\times10^{12}\,\mathrm{g\, cm^{-2}}$ was accreted. Using
the shortest suggested outburst recurrence time of $11\,\mathrm{yr}$
\citep{Degenaar12}, a superburst recurrence time of $186\,\mathrm{yr}$
is inferred, but it may very well be longer (unless the outburst
recurrence time is shorter, which would translate in shorter
superburst recurrence times).
Because of the low average luminosity, the neutron star is relatively
cool, which reduced the carbon burning rate at the bottom of the
accreted pile to allow for sufficient carbon to remain to trigger a
thermonuclear runaway after such a long recurrence time.

Of course, we cannot exclude the possibility that the superburst
ignition conditions had been almost reached during the previous
outburst, such that only a short accretion episode of a few days was
required to set it off.
Although not impossible, we find it improbable as the outer crust is
expected to have reached a higher temperature by heating during the
two-month outburst in 2000 (i.e. conditions more favorable for
ignition), than after a few days in 2011.

A more plausible scenario could be that the carbon-fuel necessary for
a superburst was created mostly during outburst, and then concentrated
during the long period of quiescence, as after accretion ceases, there
is time for the light and heavy elements to separate out from each
other \citep[see, e.g.,][]{Brown02}.
An additional potentially important process is chemical separation by
freezing at the interface of the ocean and the outer crust
\citep[see][and references therein]{Medin11}. After the previous
outburst there was plenty of time for carbon to separate out from iron
and heavier isotopes, and so substantially increasing the carbon
fraction at the bottom of the accreted column.
If this scenario is correct, then it is possible to explain the
$y_{\mathrm{ign}}$ necessary in cases where superburst ignition occurs
at early times of the outburst.
However, it could be problematic for models, as pure carbon layers
have a higher thermal conductivity and will remain colder than an
impure carbon layers \citep[see][]{Cumming01a}; moreover, upwards
transport of carbon could make it harder for the carbon to reach
ignition depth.
In any case, still unexplained is how the neutron star temperature can
rise so quickly at the start of the outburst to be able to ignite the
superburst.

The difficulties faced by superburst models that invoke carbon
ignition may point to a different fuel for superbursts. In the
analysis of bursts from the likely ultra compact X-ray binary
4U~0614+091, \citet{Kuulkers10a} pointed out that in principle helium
ignition could explain many of the observed column depths of
superbursts. This would require accumulation of a deep and cold layer
of helium on the star. Further theoretical work on this scenario is
needed, but the fact that the superburst from EXO 1745--248 appears to
have one of the largest ignition column depths of known superbursts
may place it too deep for helium ignition.

\vspace{1cm} \textbf{Acknowledgment:} We thank H. Krimm for his help
on the Swift/BAT data and J.J.M. in't Zand for insightful discussions.
RW acknowledges support from a European Research Council (ERC)
starting grant.
LK is supported by the Joint Institute for Nuclear Astrophysics (JINA;
grant PHY08-22648), a National Science Foundation Physics Frontier
Center.
ND is supported by NASA through Hubble Postdoctoral Fellowship grant
number HST-HF-51287.01-A from the Space Telescope Science Institute.
AC, GRS, and COH are supported by Discovery Grants from the Natural
Sciences and Engineering Research Council of Canada (NSERC) . COH is
also supported by an Alberta Ingenuity New Faculty Award.
JH acknowledges financial support through Chandra award GO1-12055B.
AC and LK are members of an International Team in Space Science on
type I X-ray bursts sponsored by the International Space Science
Institute (ISSI) in Bern.
DP gratefully acknowledges support provided by the National
Aeronautics and Space Administration through Chandra Award Number
GO1-12055A issued by the Chandra X-ray Observatory Center, which is
operated by the Smithsonian Astrophysical Observatory for and on
behalf of the National Aeronautics Space Administration under contract
NAS8-03060.
We thank the CXC, RXTE and Swift staff for their quick response to our
ToO requests.
This research has made use of the MAXI data provided by RIKEN, JAXA
and the MAXI team.

\end{document}